\pdfoutput=1
\documentclass[journal,comsoc]{IEEEtran}
\usepackage[T1]{fontenc}

\usepackage{cite}

\ifCLASSINFOpdf
\usepackage[pdftex]{graphicx}
\DeclareGraphicsExtensions{.eps,.pdf,.jpeg,.png}
\else
\fi
\usepackage{amsmath}
\usepackage[cmintegrals]{newtxmath}
\usepackage{eucal}

\usepackage{array,multirow}
\begin{document}
%
\title{New RLL Code with Improved Error Performance \\ for Visible Light Communication}
%
%
%

\author{Vitalio~Alfonso~Reguera,~\IEEEmembership{Member,~IEEE}
\thanks{V.A. Reguera, is with the Electrical Engineering Graduate Program, Federal University of Santa Maria, RS, Brazil   
e-mail: (vitalio.reguera@ufsm.br).}
}
\maketitle

\begin{abstract}
In this letter, a novel run-length limited (RLL) code is reported. In addition to providing a strict DC-balance and other relevant features for visible light communication (VLC) applications, the proposed 5B10B code presents enhanced error correction capabilities. Theoretical and simulation results show that the proposed code outperforms its standard enforced counterparts in terms of bit error ratio while preserving low-complexity requirements.
\end{abstract}


%
\IEEEpeerreviewmaketitle

\section{Introduction}

\IEEEPARstart{I}n recent years, there has been a growing interest in exploiting the inherent capabilities of lighting systems to transmit information. The rapid development of LED technologies has made it feasible to implement Visible Light Communication (VLC) systems to satisfy a wide range of applications \cite{Khan2017} (e.g. indoor positioning systems, Li-Fi, vehicle to vehicle communication, underwater communication, etc.). Combining lighting and communication imposes additional challenges in the design of VLC systems such as brightness control and flicker mitigation. Consequently, line codes for data transmission must be carefully chosen to avoid temporal light artefacts that could potentially harm human health.

Particularly, the IEEE standard 802.15.7 \cite{Standard2018} enforces the use of Manchester, 4B6B and 8B10B run-length limited (RLL) codes with basic modulations such as on-off key (OOK) modulation and variable pulse-position modulation (VPPM). These RLL codes provide an appropriate DC-balance to the transmitted signal at the cost of lowering the resulting data rate. Also, traditional RLL codes have weak error correction capability. Often the code rate is further affected by channel coding stages in order to improve the bit error ratio (BER). In this context, it seems convenient to join forward error correction (FEC) and RLL codes in an effort to improve error performance with a suitable computational complexity. This has been addressed in several previous works using different strategies and coding techniques \cite{Wang2015, Lu2016, Wang2016, Fang2017, Babar2017, Lu2018, Wang2019}. 

The works in \cite{Wang2015} and \cite{Wang2016} developed a soft decoding strategy for the concatenation of an outer Reed-Solomon (RS) code with an inner 4B6B or 8B10B code. In \cite{Wang2015} the RLL code is soft decoded and a hard output of one or more possible codeword candidates are passed to the RS decoder. An enhanced RLL decoding with soft output was presented in \cite{Wang2016}; improving BER when compared to its predecessors. Subsequently, the outer RS encoder was replaced by polar codes in \cite{Wang2019}. The performance of polar code is boosted using pre-determined frozen bit indexes. As a result, the proposal in \cite{Wang2019} significantly improves BER performance. These works have in common that they focus on the decoding process, but no new RLL codes were specified besides the classics 4B6B and 8B10B codes.

In \cite{Fang2017, Babar2017} new RLL encoders are considered. In \cite{Fang2017} the RLL encoder is replaced by the insertion of compensation symbols within the data frame. Unfortunately, this scheme can cause long runs of equal bits as compared to traditional RLL codes. With the purpose of increasing the capacity of the VLC systems in \cite{Babar2017}, the Unity-Rate Code (URC) was promoted to replace conventional RLL codes. This improvement in the code rate comes at the cost of relaxing the strict DC-balance at the output of the RLL encoder. For both, \cite{Fang2017} and \cite{Babar2017}, instability in the DC balance could cause flicker if brightness fluctuations exceed the maximum flickering time period (MFTP) for practical systems \cite{Rajagopal2012, Std2015, Standard2018}. 

A joint FEC-RLL coding mechanism was first introduced in \cite{Lu2016}. It combines convolutional codes and Miller codes to simultaneously achieve FEC and RLL control. A major drawback of this mechanism is the disappointing BER of Miller codes. To circumvent this problem, a new class of Miller codes, termed eMiller, was recently proposed in \cite{Lu2018}. The eMiller code match the BER of Manchester code, improving the overall performance when decoded taking advantage of their optimal trellis structure.

In this letter a novel RLL code, hereafter named 5B10B, is reported; which, in addition to preserving the desirable characteristics of traditional RLL codes for VLC, allows for enhanced error correction capabilities. Analytical and simulation results show that, for OOK-modulated transmissions over channels with moderate to high signal-to-noise ratio (SNR), the proposed 5B10B code outperforms the standard encouraged transmission techniques in terms of BER performance. 

\section{Proposed 5B10B code}

\begin{table}[!t]
\renewcommand{\arraystretch}{1.3}
	 \caption{5B10B Code}
		\label{table_1}
		\centering
		\begin{tabular}{ccc|ccc}
		\hline
		\hline
		$i$	& Data	& Code	 & $i$	 & Data 	& Code\\
		\hline
		0 &	00000	& 1100110001 & 16 &	10000 & 0111010001\\
		1 &	00001	& 1110001001 & 17 &	10001 & 0101111000\\
		2 &	00010	& 1110010010 & 18	&	10010 & 0101100011\\
		3 &	00011	& 0100011011 & 19 &	10011 & 0110101010\\
		4 &	00100	& 1101000101 & 20 &	10100 & 0110110100\\
		5 &	00101	& 1100011100 & 21 &	10101 & 0100101101\\
		6 &	00110	& 1100100110 & 22 &	10110 & 0101010110\\
		7 &	00111	& 1101001010 & 23 &	10111 & 0111001100\\
		8 &	01000	& 1001010011 & 24 &	11000 & 1001101001\\
		9 &	01001	& 1011011000 & 25 &	11001 & 0010111001\\
		10 &	01010	& 1010100011 & 26 &	11010 & 0011110010\\
		11 &	01011	& 1000111010 & 27 &	11011 & 0011001011\\
		12 &	01100	& 1001110100 & 28 &	11100 & 0011100101\\
		13 &	01101	& 1010010101 & 29 &	11101 & 0001011101\\
		14 &	01110	& 1011000110 & 30 &	11110 & 0001101110\\
		15 &	01111	& 1010101100 & 31 &	11111 & 0010011110\\
		\hline
		\end{tabular}
\end{table}

The proposed code is listed in Table \ref{table_1}. As suggested by its designation, 5B10B, maps 5-bit datawords to 10-bit codewords. Therefore, the resulting code rate is $\frac{1}{2}$; matching the code rate of Manchester code. A close inspection of Table \ref{table_1} reveals two main features of the proposed code: (\textit{i}) all codewords have the same \textit{weight}, $w=5$, and (\textit{ii}) its \textit{minimum Hamming distance} is $d_m = 4$. Thus, it can be referred to as a constant weight code with error detection and correction capabilities. The existence of such codeword subset was anticipated in \cite{Brouwer1990}, but no explicit construction was given. Here, one of these subsets was found by using computer search techniques. It is worth noting that multiple variations of the code are realizable (preserving their main features) by simply permuting the bit columns and/or by computing the ones' complement of the codewords. Moreover, according to \cite{Brouwer1990}, within the code space of constant weight, $w = 5$ and codeword length, $n = 10$, there are at least 36 codewords that satisfy $d_m = 4$.  In addition to the codewords selected, four other codewords can be included in this subset: 0000110111, 0110000111, 1000001111 and 1111100000. The reason why these codewords were not chosen to encode data bits is because they have larger runs. However, in a practical implementation they could be considered as control symbols (e.g. \textit{comma symbols} to synchronize data frames). 

Since all codewords have the same weight (with equal proportion of zeros and ones) the code guarantees a strict DC-balance. Making it similar to Manchester and 4B6B codes in this regard, and different from 8B10B which suffers from \textit{running disparity} that adds some extra complexity in the coding/decoding process. The run-length of 5B10B is six; very close to the one offered by 8B10B (five). More specifically, runs of six consecutive zeros appear when the codewords 0001011101 or 0001101110 are preceded by the 0101111000 or 1011011000.  

Most of the 5B10B properties commented so far do have aspects in common with the standard enforced RLL codes, except for the minimum distance. The proposed code doubles the minimum distance of traditional RLL codes (please, note that for Manchester, 4B6B and 8B10B codes, $d_m = 2$). A detailed analysis of the impact of this increase in minimum distance will be presented in the next subsection. It is worth emphasizing that 5B10B belongs to the group of constant weight codes. Also, other constant weight codes with different codeword length and weight could be take into account in order to increase the minimum distance. In this sense, our proposal aims to offer a balance between code rate, error correction capabilities and computational complexity. For instance, due to its relative small number of valid codewords, coding and decoding can be efficiently implemented by using a lookup table (LUT).

\subsection{Error performance}

In order to evaluate the error performance of 5B10B, in what follows we consider that VLC data is transmitted over the line-of-sight path, for which the received signal can be modeled as:

\begin{equation}\label{eq:Channelmodel}
\mathbf{y} = \alpha \mathbf{x}_{i}+\mathbf{z},
\end{equation}

where $\mathbf{x}_{i}$ represents the $i$-th OOK-modulated codeword $(i=0,2,\dots,31)$, $\alpha$ is a constant that accounts for all transmission losses and electro-optical conversion factors, and $\mathbf{z}$ is an independent and identically distributed vector of Gaussian components with zero mean and power spectral density $\frac{N_0}{2}$, representing both thermal noise and background shot noise. Assuming that signal symbols are equiprobable, the \textit{maximum likelihood} (ML) detector for this channel model is given by:

\begin{equation}\label{eq:ML}
\hat{i} =  \underset{0 \leq i \leq 31}{\text{arg min}} \Vert \mathbf{y} - \mathbf{x}_i \Vert,
\end{equation}
 
and the resulting symbol error probability is upper bounded by the union bound as:

\begin{equation}\label{eq:Pe1}
P_e \leq \frac{1}{32} \sum_{r=1}^{5} \sum_{i=0}^{31} A_{i,r} Q\left( \sqrt{\frac{r \Delta^2}{2 N_0}} \right), 
\end{equation}

where $A_{i,r}$ is the number of codewords that are at Hamming distance $2r$ from the $i$-th codeword (please note that since it is a constant weight code, the Hamming distance between any two codewords is a multiple of 2), $Q(\cdot)$ is the complementary cumulative distribution function of the standard normal distribution, and $\Delta$ is the Euclidean distance between two OOK-modulated signal symbol that differ in the position of just one ON pulse. Defining $\mathcal{E}$ as the energy of the difference between individually received ON and OFF pulses, we have:

\begin{equation}\label{eq:D}
\Delta = \sqrt{2\mathcal{E}} 
\end{equation}

\begin{table}[!t]
\renewcommand{\arraystretch}{1.3}
	 \caption{Average number of codewords at Hamming distance $2r$ over the entire codeword set: $\frac{1}{32}\sum_{i=0}^{31} A_{i,r}$}
		\label{table_2} 
		\centering
		\begin{tabular}{ccccc}
		\hline
		\hline
		$r=1$	& $r=2$	 & $r=3$ 	& $r=4$ & $r=5$\\
		\hline
		0	& 17.69 & 8.81 & 4.50 & 0\\
		\hline
		\end{tabular}
\end{table}

Table \ref{table_2} shows the average number of codewords per symbol at Hamming distance $2r$. Since 5B10B has $d_m = 4$, $A_{i,1} = 0, \forall i$. It is also easy to verify from Table \ref{table_1} that, for any codeword, its one's complement does not produce a valid codeword. Therefore, $A_{i,5} = 0, \forall i$. Using Equation (\ref{eq:D}) and the results in Table \ref{table_2}, Equation (\ref{eq:Pe1}) can be rewritten as:

\begin{equation}\label{eq:Pe2}
P_e \leq 17.69 \cdot Q\left( \sqrt{\frac{2 \mathcal{E}}{N_0}} \right) +  8.81 \cdot Q\left( \sqrt{\frac{3 \mathcal{E}}{N_0}} \right) +  4.5 \cdot Q\left( \sqrt{\frac{4 \mathcal{E}}{N_0}} \right)
\end{equation}

For the proposed code, it's not hard to see that the bit error probability, $P_b$, will be bounded as:

\begin{equation}\label{eq:Pb_Bound}
\frac{P_e}{5} < P_b < P_e.
\end{equation}

In order to bring the actual $P_b$ as close as possible to its theoretical lower bound, the \textit{binary switching algorithm} (BSA) \cite{Zeger1990} was used to efficiently map datawords to codewords (as presented in Table \ref{table_1}). The algorithm is started with an arbitrary assignment of binary labels to codewords. A cost function, $\Phi(i)$, was used to weigh the contribution of each pair-wise assignment to the bit error probability as follows:

\begin{equation}\label{eq:C}
\Phi(i) = \frac{1}{32} \sum_{\forall j \neq i} d_{i,j} Q\left(\sqrt{\frac{\Delta_{i,j}^{2}}{2 N_0}} \right),
\end{equation}

where $d_{i,j}$ is the Hamming distance between the datawords assigned to the $i$-th and $j$-th codewords $(i,j = \{0,1,2\dots,31\})$, and $\Delta_{i,j}$ is the Euclidean distance between the same pair $(i,j)$ of OOK-modulated codewords. The algorithm minimizes the overall cost function iteratively by swapping labels between codewords in a pair-wise fashion. The swapping operation  start with those codewords that present a greater contribution to $P_b$. The procedure continues until the label swapping operation does not allow any further reduction of the overall cost function. For any given mapping, the overall cost function is computed as $\sum_{i=0}^{31} \Phi(i)$. For large SNR, this function provides a good estimate of the average number of bits received with errors per transmitted symbol. From where, $P_b$ can be approximated as:

\begin{equation}\label{eq:Pb}
P_b \sim \frac{1}{5} \sum_{i=0}^{31} \Phi(i).
\end{equation}

It is important to point out that, depending on the initial binary assignment, the BSA strategy can lead to different final mappings with very similar expected overall error performance. Thus, alternative mappings can be considered to meet specific application demands. Also, other cost functions that reflect a different system model can be easily adopted if required.

\section{Simulation Results}

\begin{figure}[!t]
\centering
\includegraphics[width=0.85\columnwidth]{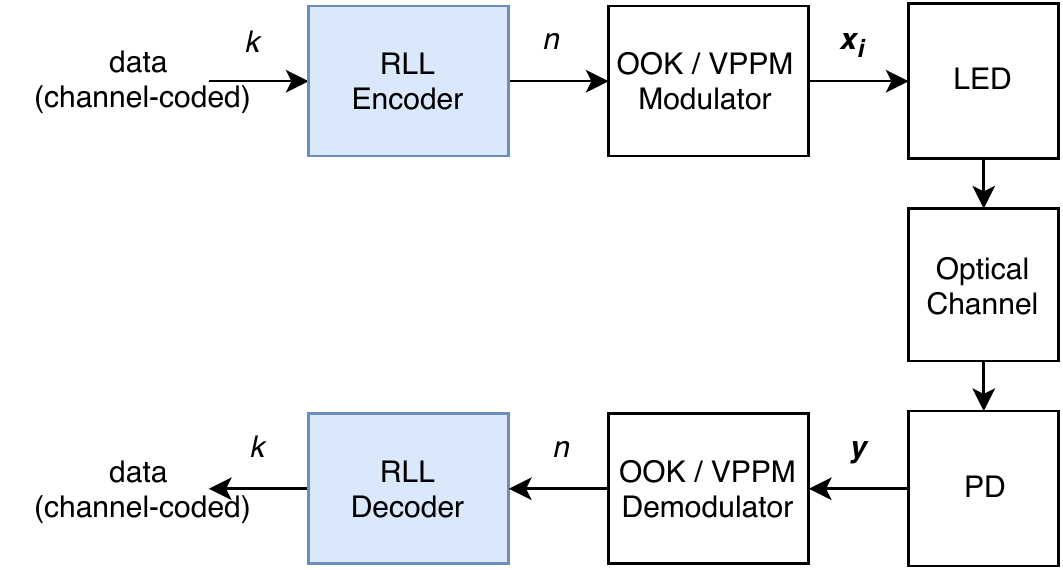}
\caption{Block diagram of the VLC system model} 
\label{fig:System}
\end{figure}

For the sake of validating the proposed code, in this section, the results of extensive simulations are presented. The system model used in the simulations is depicted in Fig. \ref{fig:System}. The RLL encoder maps $k$-bit datawords to $n$-bit codewords that are subsequently modulated and transmitted over the optical channel. The optical signal is captured by a photo-detector (PD) at the receiver and the correspondent decoding process is performed. Code 5B10B was contrasted to physical layer RLL codes specified for PHY I and PHY II in \cite{Standard2018}. As recommended, the 4B6B code was modulated using VPPM; with a 50\% duty cycle. While Manchester and 8B10B RLL codes, as well as the proposed 5B10B code, were OOK-modulated. The optical channel was modeled as described by Equation (\ref{eq:Channelmodel}). In the experiments, the transmission of $10^8$ codewords carrying a random pattern of data bits was considered in calculating the \textit{power spectral density} (PSD), the \textit{symbol error ratio} (SER) and the BER. 

\begin{figure}[!t]
\centering
\includegraphics[width=0.95\columnwidth]{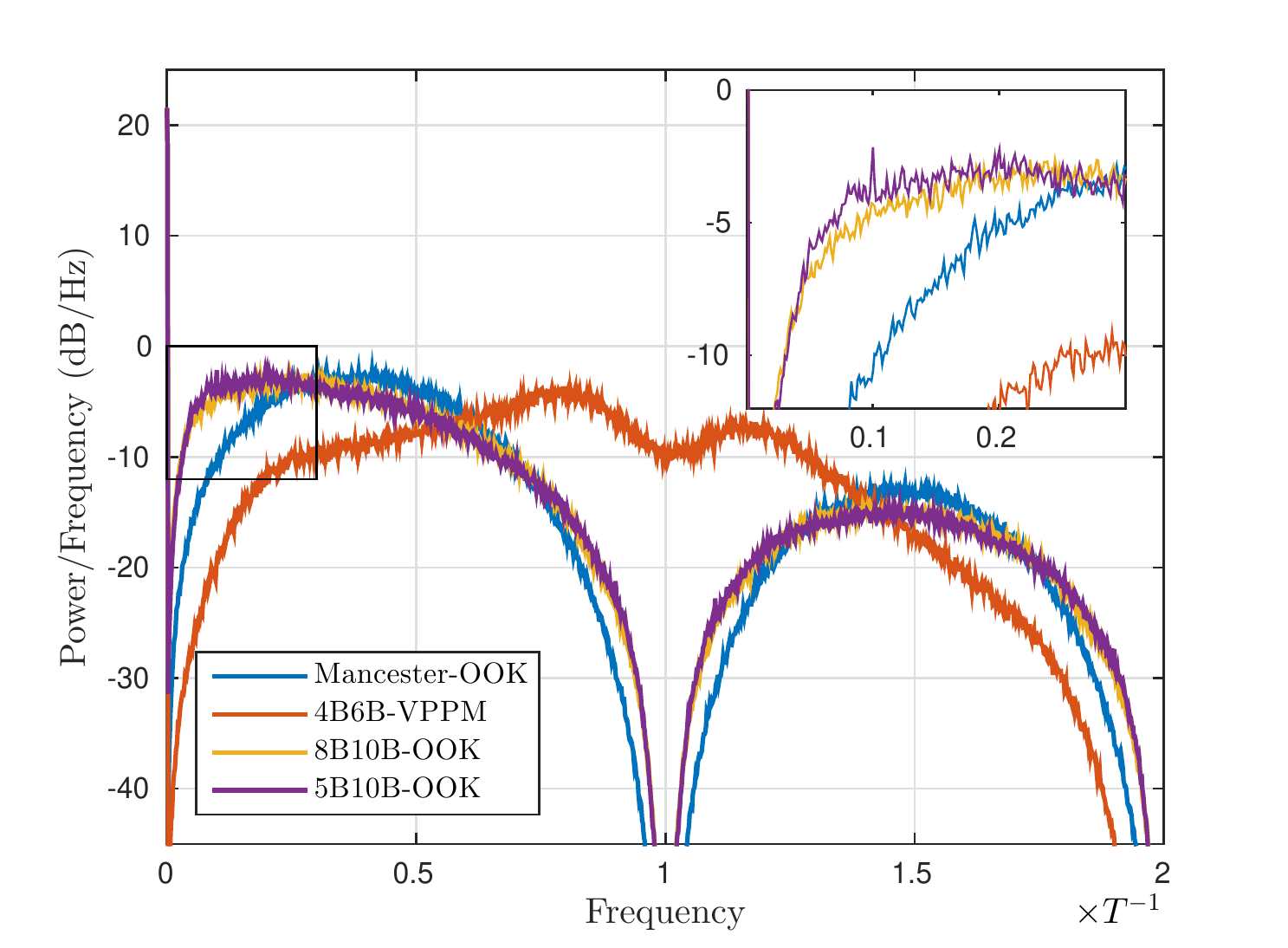}
\caption{Power Spectral Density estimates, comparison between standardize RLL codes and the proposed 5B10B code.} 
\label{fig:PSD}
\end{figure}

Fig \ref{fig:PSD} shows the PSD for the analysed RLL codes. The frequency scale is normalized by the inverse of the pulse period ($T^{-1}$). As expected, the spectral content of 5B10B is very close to that produced by the 8B10B encoder. The slight increase in the run-length, makes the proposed code to shift a small portion of the spectral components towards low frequencies, when compared to 8B10B. This marginal increase in low frequency spectral components can be perceived in more detail in the zoomed box on the upper right side of Fig \ref{fig:PSD}. Except for the DC component, there is a sharp decrease in spectral components below $0.05 \times T^{-1}$Hz. Taking as reference the minimum clock rate specified in \cite{Standard2018}, noticeable oscillations will appear at frequencies greater than $0.05 \times 200\text{KHz}=10\text{KHz}$. This guarantees that the MFTP stays well below the recommended values to minimize health risks associated with low-frequency modulation of lighting sources \cite{Std2015, Standard2018}.

\begin{figure}[!t]
\centering
\includegraphics[width=0.95\columnwidth]{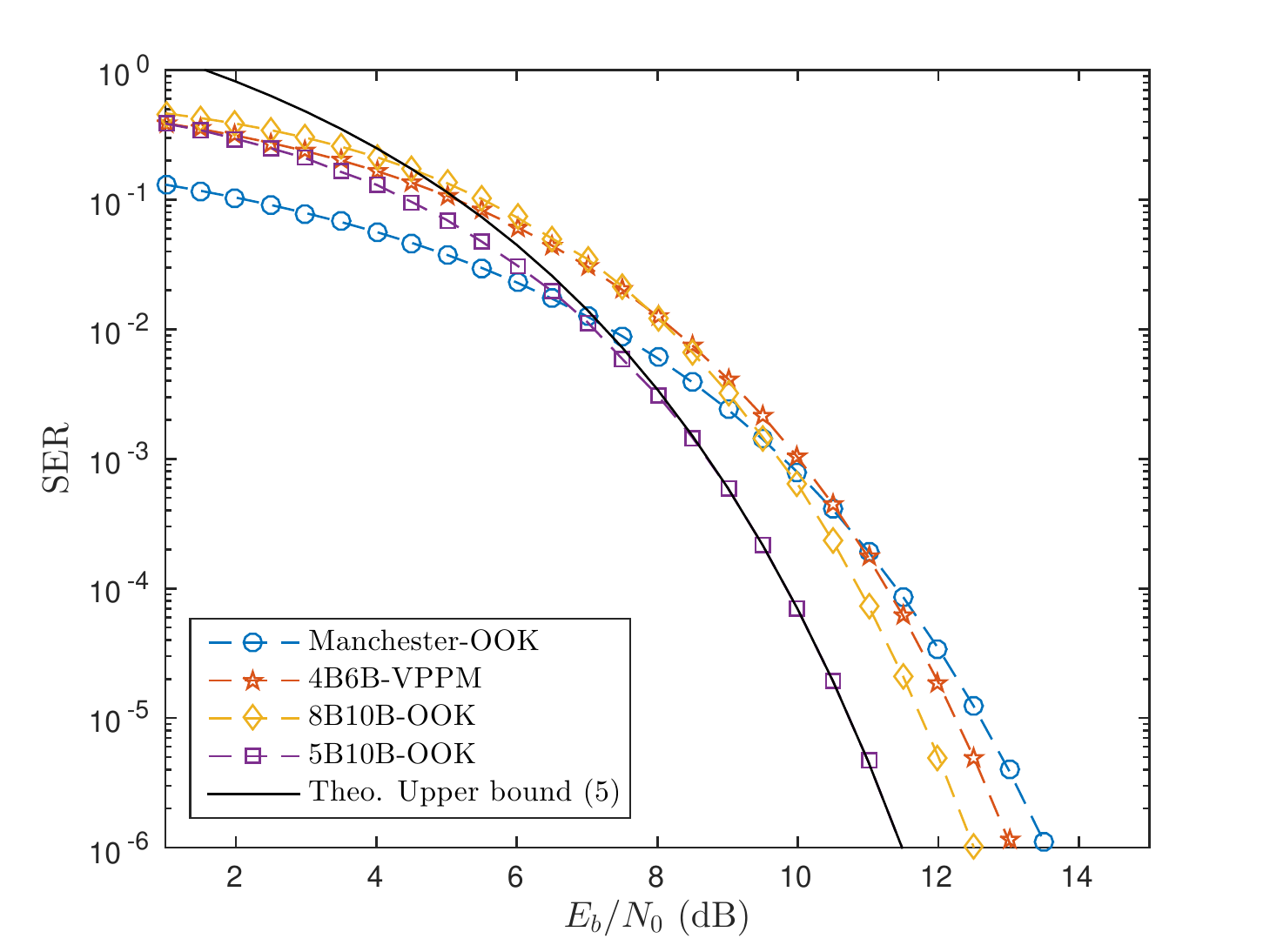}
\caption{SER curves for different RLL codes, including the observed and theoretically estimated values for the proposed 5B10B code.} 
\label{fig:SER}
\end{figure}

The error performance of the 5B10B code is shown in Fig \ref{fig:SER}. The upper bound of $P_e$, calculated from Equation (\ref{eq:Pe2}), is depicted in Fig. \ref{fig:SER} with a dark solid line. As the SNR increases, the union bound becomes a tight bound of the symbol error probability. Moreover, for values of the energy per bit to noise power spectral density ratio, $\frac{E_b}{N_0}$, above $9$dB, Equation (\ref{eq:Pe2}) offers a good estimate of the actual SER. For $\frac{E_b}{N_0}$ greater than $6.79$dB, the proposed code outperforms Manchester, 4B6B and 8B10B encoders. Particularly, compared to Manchester coding, 5B10B requires approximately $2$dB less of signal energy to offer a SER $=10^{-6}$. It should be noted that in all cases the error performance of the system can be improved by using additional channel coding techniques (as suggested in Fig. \ref{fig:System}) at the cost of lowering the data rate. For instance, the output of a RS encoder can be coupled to the input of the RLL encoder. This follows trivially from the well-known result on concatenated codes, \cite{Forney1965}, that the concatenation of an outer code with 5B10B has the potential to double the error correction capabilities compared to its standardized counterparts. However, to achieve this potential improvement, proper selection of the algorithms and coding parameters is required. Also, decoding algorithms such as those proposed in \cite{Wang2015} and \cite{Wang2016} can be implemented together with 5B10B in other to improve the overall BER.

Table \ref{table_3} summarizes the main characteristics of the proposed RLL code and establishes a comparison with its counterparts.  As discussed above, it resembles in some aspects the RLL codes commonly used for VLC. The data rate that can be achieved with 5B10B is equal to that of the Manchester code, while its run-length is very close to that of the 8B10B code. However, differently from standardized RLL codes, it has the ability not only to detect but also to correct errors. Last column of Table \ref{table_3} shows the $\frac{E_b}{N_0}$ required by the RLL codes under study at BER = $10^{-5}$. Since the error performance for an OOK-modulated Manchester encoded transmission is the same as that of uncoded OOK or VPPM (with 50\% duty cycle) \cite{Abshire1984}, its energy requirement can be used as benchmark to estimate the coding gain. Hence, it follows that at BER = $10^{-5}$ the coding gain offered by Manchester, 4B6B, 8B10B and 5B10B is: 0dB, 0.43dB, 0.86dB and 2.17dB, respectively. It is worth noting that the achieved coding gain is comparable to those reported in \cite{Wang2015} and \cite{Lu2018}. However, in the cited works this output comes at the cost of a significant increase in computational complexity when compared to the current proposal.

\begin{table}[!t]
\renewcommand{\arraystretch}{1.1}
	 \caption{Performance comparison between 5B10B and standard enforced RLL codes for VLC}
		\label{table_3} 
		\centering
		\begin{tabular}{ccccc}
		\hline
		\multirow{2}{*}{RLL} & \multirow{2}{*}{Modulation} & Run- &  Spectral & $E_b/N_0$\\
		& & length & efficiency & (BER = $10^{-5}$) \\
		\hline
		Manchester	& OOK 	& 2 & 0.5 bit/s/Hz & 12.59 dB\\
		4B6B		& VPPM 	& 4 & $0.\bar{3}$ bit/s/Hz & 12.16 dB\\
		8B10B		& OOK 	& 5 & 0.8 bit/s/Hz & 11.73 dB\\
		5B10B		& OOK 	& 6 & 0.5 bit/s/Hz & 10.42 dB\\
		\hline
		\end{tabular}
\end{table}

\section{Conclusion}
The proposed 5B10B code is analogous in many aspects to the RLL codes commonly used in VLC applications. Particularly, it ensures strict DC-balance, run-length limited to six consecutive equal bits and $\frac{1}{2}$ code rate. However, different from its standardized counterparts, it offers enhanced error correction capabilities. The main advantage of the proposed 5B10B RLL code is that it provides an extra coding gain (2.17dB at BER = $10^{-5}$, based on theoretical and simulation results) without incurring an increase in computational complexity, making it a promising candidate for VLC applications.

\ifCLASSOPTIONcaptionsoff
  \newpage
\fi

\bibliographystyle{IEEEtran}
\bibliography{VLC_paper}

\begin{thebibliography}{10}
\providecommand{\url}[1]{#1}
\csname url@samestyle\endcsname
\providecommand{\newblock}{\relax}
\providecommand{\bibinfo}[2]{#2}
\providecommand{\BIBentrySTDinterwordspacing}{\spaceskip=0pt\relax}
\providecommand{\BIBentryALTinterwordstretchfactor}{4}
\providecommand{\BIBentryALTinterwordspacing}{\spaceskip=\fontdimen2\font plus
\BIBentryALTinterwordstretchfactor\fontdimen3\font minus
  \fontdimen4\font\relax}
\providecommand{\BIBforeignlanguage}[2]{{%
\expandafter\ifx\csname l@#1\endcsname\relax
\typeout{** WARNING: IEEEtran.bst: No hyphenation pattern has been}%
\typeout{** loaded for the language `#1'. Using the pattern for}%
\typeout{** the default language instead.}%
\else
\language=\csname l@#1\endcsname
\fi
#2}}
\providecommand{\BIBdecl}{\relax}
\BIBdecl

\bibitem{Khan2017}
L.~U. Khan, ``Visible light communication: Applications, architecture,
  standardization and research challenges,'' \emph{Digital Communications and
  Networks}, vol.~3, no.~2, pp. 78 -- 88, 2017.

\bibitem{Standard2018}
``{IEEE Approved Draft Standard for Local and metropolitan area networks - Part
  15.7: Short-Range Optical Wireless Communications},'' \emph{IEEE
  P802.15.7/D3a, August 2018}, pp. 1--428, Jan 2018.

\bibitem{Wang2015}
H.~{Wang} and S.~{Kim}, ``{New RLL Decoding Algorithm for Multiple Candidates
  in Visible Light Communication},'' \emph{IEEE Photonics Technology Letters},
  vol.~27, no.~1, pp. 15--17, Jan 2015.

\bibitem{Lu2016}
X.~{Lu} and J.~{Li Tiffany}, ``{Achieving FEC and RLL for VLC: A Concatenated
  Convolutional-Miller Coding Mechanism},'' \emph{IEEE Photonics Technology
  Letters}, vol.~28, no.~9, pp. 1030--1033, May 2016.

\bibitem{Wang2016}
H.~{Wang} and S.~{Kim}, ``{Soft-Input Soft-Output Run-Length Limited Decoding
  for Visible Light Communication},'' \emph{IEEE Photonics Technology Letters},
  vol.~28, no.~3, pp. 225--228, Feb 2016.

\bibitem{Fang2017}
J.~{Fang}, Z.~{Che}, Z.~L. {Jiang}, X.~{Yu}, S.~{Yiu}, K.~{Ren}, X.~{Tan}, and
  Z.~{Chen}, ``{An Efficient Flicker-Free FEC Coding Scheme for Dimmable
  Visible Light Communication Based on Polar Codes},'' \emph{IEEE Photonics
  Journal}, vol.~9, no.~3, pp. 1--10, June 2017.

\bibitem{Babar2017}
Z.~{Babar}, H.~V. {Nguyen}, P.~{Botsinis}, D.~{Alanis}, D.~{Chandra}, S.~X.
  {Ng}, and L.~{Hanzo}, ``{Unity-Rate Codes Maximize the Normalized Throughput
  of On-Off Keying Visible Light Communication},'' \emph{IEEE Photonics
  Technology Letters}, vol.~29, no.~3, pp. 291--294, Feb 2017.

\bibitem{Lu2018}
X.~{Lu} and J.~{Li}, ``{New Miller Codes for Run-Length Control in Visible
  Light Communications},'' \emph{IEEE Transactions on Wireless Communications},
  vol.~17, no.~3, pp. 1798--1810, March 2018.

\bibitem{Wang2019}
H.~{Wang} and S.~{Kim}, ``{Design of Polar Codes for Run-Length Limited Codes
  in Visible Light Communications},'' \emph{IEEE Photonics Technology Letters},
  vol.~31, no.~1, pp. 27--30, Jan 2019.

\bibitem{Rajagopal2012}
S.~{Rajagopal}, R.~D. {Roberts}, and S.~{Lim}, ``{IEEE 802.15.7 visible light
  communication: modulation schemes and dimming support},'' \emph{IEEE
  Communications Magazine}, vol.~50, no.~3, pp. 72--82, March 2012.

\bibitem{Std2015}
``{IEEE Recommended Practices for Modulating Current in High-Brightness LEDs
  for Mitigating Health Risks to Viewers},'' \emph{IEEE Std 1789-2015}, pp.
  1--80, June 2015.

\bibitem{Brouwer1990}
A.~E. {Brouwer}, J.~B. {Shearer}, N.~J.~A. {Sloane}, and W.~D. {Smith}, ``{A
  new table of constant weight codes},'' \emph{IEEE Transactions on Information
  Theory}, vol.~36, no.~6, pp. 1334--1380, Nov 1990.

\bibitem{Zeger1990}
K.~{Zeger} and A.~{Gersho}, ``{Pseudo-Gray coding},'' \emph{IEEE Transactions
  on Communications}, vol.~38, no.~12, pp. 2147--2158, Dec 1990.

\bibitem{Forney1965}
G.~D. Forney~Jr, ``{Concatenated codes.}'' Ph.D. dissertation, Massachusetts
  Institute of Technology, 1965.

\bibitem{Abshire1984}
J.~{Abshire}, ``{Performance of OOK and Low-Order PPM Modulations in Optical
  Communications When Using APD-Based Receivers},'' \emph{IEEE Transactions on
  Communications}, vol.~32, no.~10, pp. 1140--1143, October 1984.

\end{thebibliography}

\end{document}